\documentclass[paper=A4,11pt,DIV=12,headings=big]{scrartcl}
\pdfoutput=1

\usepackage[a4paper, top=1.2in, bottom=1.2in, left=1.1in, right=1.1in]{geometry}
\usepackage{amsfonts}
\usepackage{amsmath,amssymb}
\usepackage{mathrsfs}
\usepackage[normal]{caption}

\usepackage{cancel}
\usepackage{xcolor}
\usepackage{cite,bbm}
\usepackage{enumerate}
\usepackage{graphicx}
\graphicspath{{./Figs/}}
\usepackage{wrapfig}

\renewenvironment{abstract}{%
\begin{minipage}{0.95\textwidth}
}
{\par\noindent\end{minipage}}

\usepackage[multiple]{footmisc}
\let\oldfootnote\footnote\renewcommand\footnote[1]{\oldfootnote{\hspace{2mm}#1}}

\definecolor{darkblue}{rgb}{0,0,0.9}
\usepackage{hyperref}
\hypersetup{
linktocpage,
colorlinks,
citecolor=darkblue,
filecolor=darkblue,
linkcolor=darkblue,
urlcolor=darkblue
}

\allowdisplaybreaks
\addtokomafont{disposition}{\rmfamily\boldmath}

\newcommand{\mc}{\mathcal}
\renewcommand{\th}{{\tilde h}}

\newcommand{\<}{\langle}
\renewcommand{\>}{\rangle}

\newcommand{\re}{{\rm Re}}

\newcommand{\alem}{\alpha_{\rm em}}
\newcommand{\sh}{\hat s}
\renewcommand{\th}{\hat t}
\newcommand{\uh}{\hat u}
\newcommand{\mh}{\hat m}
\newcommand{\Mh}{\hat M}

\def\sla#1{\setbox0=\hbox{$#1$}\dimen0=\wd0
      \setbox1=\hbox{/} \dimen1=\wd1 \ifdim\dimen0>\dimen1
      \rlap{\hbox to \dimen0{\hfil/\hfil}} #1                        \else
      \rlap{\hbox to \dimen1{\hfil$#1$\hfil}}
      /   \fi}

\newcommand{\be}{\begin{equation}}
\newcommand{\ee}{\end{equation}}
\newcommand{\bea}{\begin{eqnarray}}
\newcommand{\eea}{\end{eqnarray}}

\newcommand{\nn}{\nonumber}

\begin{document}

\begin{flushright}
\small
LAPTH-023/16
\end{flushright}
\vskip0.5cm

\begin{center}
{\sffamily \bfseries \LARGE \boldmath
More Lepton Flavor Violating Observables\\[0.1cm]for LHCb's Run 2}\\[0.8 cm]
{\normalsize \sffamily \bfseries Diego Guadagnoli$^a$, Dmitri Melikhov$^{b,c}$ and M\'eril Reboud$^{a,d}$} \\[0.5 cm]
\small
$^a${\em Laboratoire d'Annecy-le-Vieux de Physique Th\'eorique UMR5108\,, Universit\'e de Savoie Mont-Blanc et CNRS, B.P.~110, F-74941, Annecy-le-Vieux Cedex, France}\\[0.1cm]
$^b${\em Institute of Nuclear Physics, Moscow State University, 119992, Moscow, Russia}\\[0.1cm]
$^c${\em Faculty of Physics, University of Vienna, Boltzmanngasse 5, A-1090 Vienna, Austria}
\\[0.1cm]
$^d${\em \'Ecole Normale Sup\'erieure de Lyon, 46 all\'ee d'Italie, F-69364, Lyon Cedex 07, France}
\end{center}

\medskip

\begin{abstract}\noindent
The $R_K$ measurement by LHCb suggests non-standard lepton non-universality (LNU) to occur in $b \to s \ell^+ \ell^-$ transitions, with effects in muons rather than electrons. A number of other measurements of $b \to s \ell^+ \ell^-$ transitions by LHCb and $B$-factories display disagreement with the SM predictions and, remarkably, these discrepancies are consistent in magnitude and sign with the $R_K$ effect. Non-standard LNU suggests non-standard lepton flavor violation (LFV) as well, for example in $B \to K \ell \ell'$ and $B_s \to \ell \ell'$. There are good reasons to expect that the new effects may be larger for generations closer to the third one. In this case, the $B_s \to \mu e$ decay may be the most difficult to reach experimentally. We propose and study in detail the radiative counterpart of this decay, namely $B_s \to \mu e \gamma$, whereby the chiral-suppression factor is replaced by a factor of order $\alpha / \pi$. A measurement of this mode would be sensitive to the same physics as the purely leptonic LFV decay and, depending on experimental efficiencies, it may be more accessible. A realistic expectation is a factor of two improvement in statistics for either of the $B_{d,s}$ modes.
\end{abstract}

\vspace{1.0cm}

\renewcommand{\thefootnote}{\arabic{footnote}}
\setcounter{footnote}{0}

\noindent {\bf Introduction --} During run 1, the LHCb experiment performed a number of  measurements of $b \to s$ and $b \to c$ transitions, finding less than perfect agreement with the Standard Model (SM). On the same timescale, updates on some of the most interesting of these measurements were published by the $B$ factories as well, with results consistent with LHCb's.

In more detail, the experimental situation can be summarized as follows. The most striking effect is that measured by LHCb in the ratio known as $R_{K}$\cite{Aaij:2014ora}
\be\label{eq:RK}
R_K \equiv \frac{\mc B (B^+ \to K^+ \mu^+\mu^-)}{\mc B (B^+ \to K^+ e^+e^-)} =
0.745^{+0.090}_{-0.074}\,{\rm (stat)}\pm 0.036\,{\rm (syst)}\,,
\ee
in the di-lepton invariant-mass-squared range $[1, 6]$ GeV$^{2}$. The SM predicts unity with percent-level corrections \cite{Bordone:2016gaq,Bobeth:2007dw,Bouchard:2013mia,Hiller:2003js}, implying a 2.6$\sigma$ discrepancy. Two convincing aspects of this discrepancy are the theoretical cleanness of $R_{K}$, and the fact that the measurement in the electron channel, the one more subject to large systematics, is, within errors, in agreement with the SM prediction \cite{Aaij:2014ora}. On the other hand, the muon-channel measurement, expected to be experimentally more solid, yields~\cite{Aaij:2014pli,Aaij:2012vr}
\be\label{eq:NewBRKmumu}
\mc B(B^+ \to K^+\mu^+\mu^-)_{[1,6]} = (1.19 \pm 0.03 \pm 0.06)\times 10^{-7}~,
\ee
which is about $30\%$ lower than the SM prediction, $\mc B(B^+ \to K^+ \mu^+ \mu^-)^{\rm SM}_{[1,6]} = (1.75^{+0.60}_{-0.29}) \times 10^{-7}$~\cite{Bobeth:2011gi,Bobeth:2011nj,Bobeth:2012vn}.

The very same pattern, with data lower than the SM prediction, is actually also observed in the $B_{s} \rightarrow \phi \mu^{+} \mu^{-}$ channel and in the same range $m_{\mu \mu}^{2} \in [1, 6]$ GeV$^{2}$, as initially found in 1/fb of LHCb data \cite{Aaij:2013aln} and then confirmed by a full run-1 analysis \cite{Aaij:2015esa}. This discrepancy is estimated to be more than 3$\sigma$ \cite{Aaij:2015esa}.

Additional support comes from the $B \rightarrow K^{*} \mu \mu$ decay, for which LHCb can perform a full angular analysis. The quantity known as $P_{5}'$, designed to have reduced sensitivity to form-factor uncertainties \cite{Descotes-Genon:2013vna}, exhibits a discrepancy in two bins, again in the low-$m^{2}_{\mu\mu}$ range. The effect was originally found in 1/fb of LHCb data \cite{Aaij:2013qta}, and confirmed by a full run-1 analysis \cite{Aaij:2015oid} as well as, very recently, by a Belle analysis  \cite{Abdesselam:2016llu}. The $P_{5}'$ discrepancy as estimated by LHCb amounts to $3.4\sigma$, and is in the 2$\sigma$-ballpark from Belle (2.1$\sigma$ as compared to \cite{Descotes-Genon:2014uoa} and 1.7$\sigma$ as compared to \cite{Straub:2015ica,Jager:2012uw,Jager:2014rwa}). The theoretical error is, however, still debated, see in particular \cite{Descotes-Genon:2013wba,Lyon:2014hpa,Jager:2014rwa,Ciuchini:2015qxb}.

Further interesting results come from measurements of the ratios $R(D^{(*)}) \equiv \mc B (B \to D^{(*)} \tau \nu) / \mc B (B \to D^{(*)} \ell \nu)$. They were initially reported by BaBar \cite{Lees:2012xj} to be in excess of the SM prediction \cite{Fajfer:2012vx,Kamenik:2008tj}. The excess in the $R(D^{*})$ channel was recently confirmed by LHCb in 3/fb of run-1 data \cite{Aaij:2015yra}. Consistent results were also reported by Belle in two analyses, using respectively hadronically- \cite{Huschle:2015rga} and semileptonically-decaying \cite{Belle-semilep} taus.

On the theoretical side, it is likewise remarkable that a consistent picture of all
the above-mentioned effects is possible already within an effective-theory approach. Global fits to the Wilson coefficients of the general $b \rightarrow s$ effective Hamiltonian point towards new-physics (NP) shifts either in $C_9$ only, or in the $SU(2)_L$-invariant direction $C_9^{\rm NP} = - C_{10}^{\rm NP}$, with comparable $\chi^2$ between the two cases \cite{Ghosh:2014awa,Altmannshofer:2014rta}. (See also \cite{Beaujean:2013soa,Hurth:2014vma,Beaujean:2015gba,Descotes-Genon:2015uva,Hurth:2016fbr}.) The corresponding terms in the effective Hamiltonian read
\be
\mathscr{H}_{\rm SM + NP}(\bar b \to \bar s \ell^+ \ell^-) ~=~ 
- \frac{4 G_F}{\sqrt 2} V^*_{tb} V_{ts} \frac{\alem (m_b)}{4 \pi} \, \bar b_L \gamma^\lambda s_L \, \bar \ell (\,C_9^{\ell} \, \gamma_\lambda + C_{10}^{\ell} \, \gamma_\lambda \gamma_5) \ell + {\rm H.c.},
\ee
where $C_{9,10}^{\ell} = C_{9,10}^{\rm SM} + C_{9,10}^{{\rm NP}, \ell}$. The $\ell$ label takes into account that the NP contributions to the Wilson coefficients may depend on the lepton flavor. This possibility, suggested by the $R_K$ result and referred to as lepton-flavor non-universality (LNU), has inspired a number of SM extensions where new LNU interactions are introduced via the exchange of multi-TeV particles \cite{Altmannshofer:2013foa,Gauld:2013qba,Buras:2013qja,Gauld:2013qja,Buras:2013dea,Altmannshofer:2014cfa,Biswas:2014gga,Hiller:2014yaa,Ghosh:2014awa,Glashow:2014iga,Hiller:2014ula,Gripaios:2014tna,Crivellin:2015mga,Crivellin:2015lwa,Niehoff:2015bfa,Sierra:2015fma,Celis:2015ara,Becirevic:2015asa,Carmona:2015ena,Celis:2015eqs,Varzielas:2015iva,Lee:2015qra,Boucenna:2015raa,Crivellin:2015era,Belanger:2015nma,Alonso:2015sja,Greljo:2015mma,Altmannshofer:2015mqa,Guadagnoli:2015nra,Calibbi:2015kma,Sahoo:2015pzk,Falkowski:2015zwa,Buttazzo:2016kid,Chiang:2016qov,Belanger:2016ywb,Becirevic:2016zri}. 

If data confirm the presence of beyond-SM LNU, in general -- namely, in the absence of further assumptions -- we also expect non-standard LFV \cite{Glashow:2014iga}.\footnote{%
Forbidding non-standard LFV within models able to explain $R_K$ requires a dynamical or a symmetry mechanism, that for example extends the SM lepton-flavor symmetries to the new model. Attempts in this direction are in refs. \cite{Celis:2015ara,Alonso:2015sja}.} One may object that this is not the case already in the SM plus any minimal mechanism for neutrino masses, as in this case LFV in decay is suppressed by $m_\nu$. However, there is no compelling argument why the new physics explaining $R_K$ should have neutrino masses as the only LFV spurions.

More specifically, ref. \cite{Glashow:2014iga} proposed that the $R_K$ deviation from unity be due to an effective interaction involving dominantly quarks and leptons of the third generation, namely
\be
\label{eq:HNP}
\mathscr{H}_{\rm NP} ~=~ G \, \bar{b}'_L \gamma^\lambda b'_L 
\bar{\tau}'_L \gamma_\lambda \tau'_L~.
\ee
If the scale of this interaction is above the electroweak-symmetry breaking one, quarks and leptons are, in general, in the gauge basis, indicated with a prime. After rotation to the mass eigenbasis, one can then expect effects that are largest for third-generation quarks and leptons, and suppressed accordingly for lighter generations.\footnote{%
Actually, given the O(TeV)-scale of the new interactions, it is expected that the fermionic d.o.f. involved be complete multiplets under the unbroken EW symmetry \cite{Alonso:2014csa}. This observation establishes correlations \cite{Bhattacharya:2014wla} between (among the others) effects in $b \to s$ and in $b \to c$ transitions, thus allowing a common origin for the $b \to s \ell \ell$ discrepancies and those in $R(D)$ and $R(D^{(*)})$. For further quantitative studies see \cite{Bauer:2015knc,Fajfer:2015ycq,Calibbi:2015kma,Greljo:2015mma,Buttazzo:2016kid,Boucenna:2016wpr}.}

Assuming the interaction (\ref{eq:HNP}), the amount of LNU pointed to by $R_K$ actually allows to quantify rather generally \cite{Glashow:2014iga} the expected amount of LFV. In fact, $R_K$ yields the ratio
\be
\rho_{\rm NP} = -0.159^{+0.060}_{-0.070}
\ee
between the NP and the SM+NP contribution to $C_9^\mu$. Then
\be
\label{eq:BKll}
\frac{\mc B (B \to K \ell_i^\pm \ell_j^\mp)}{\mc B (B^+ \to K^+ \mu^+ \mu^-)} ~\simeq~ 
2 \rho_{\rm NP}^2 
\frac{| (U^\ell_L)_{3i} |^2 | (U^\ell_L)_{3j} |^2}{|(U^\ell_L)_{32}|^4}~,
\ee
implying
\bea
\label{eq:BKlilj}
\mc B (B \to K \ell_i^\pm \ell_j^\mp) &\simeq& 5\% \, \cdot \, 
\mc B (B^+ \to K^+ \mu^+ \mu^-) \, \cdot \,  
\frac{| (U^\ell_L)_{3i} |^2 | (U^\ell_L)_{3j} |^2}{|(U^\ell_L)_{32}|^4} \nn \\
&\simeq& 2.2 \times 10^{-8} \, \cdot \,
\frac{| (U^\ell_L)_{3i} |^2 | (U^\ell_L)_{3j} |^2}{|(U^\ell_L)_{32}|^4}~,
\eea
where we used $\mc B (B^+ \to K^+ \mu^+ \mu^-) \simeq 4.3 \times 10^{-7}$ \cite{Aaij:2014pli}, and neglected all terms proportional to the different masses of the final-state leptons.\footnote{Because of this approximation, eqs. (\ref{eq:BKll})-(\ref{eq:BKlilj}) provide only crude estimates in the case of decays involving a $\tau$ lepton. However, this approximation does not change the argument of the present paragraph.} Eq. (\ref{eq:BKlilj}) tells us that LFV $B \to K$ decays are expected to be in the ballpark of $10^{-8}$ times an unknown factor involving $U_L^\ell$ matrix entries. In the $\ell_i \ell_j = e \mu$ case, this ratio reads $|(U_L^\ell)_{31} / (U_L^\ell)_{32}| \lesssim 3.7$ \cite{Glashow:2014iga}, implying that the $B \to K \mu e$ rate may be around $10^{-8}$, or much less if $|(U_L^\ell)_{31} / (U_L^\ell)_{32}| \ll 1$. The latter possibility would suggest $U_L^\ell$ entries that decrease in magnitude with the distance from the diagonal. But then one may expect the ratio $|(U_L^\ell)_{33} / (U_L^\ell)_{32}| > 1$, implying a $B \to K \mu \tau$ rate of O($10^{-8}$) or above. In short, assuming the interaction (\ref{eq:HNP}), one can hope that at least one LFV $B \to K$ decay rate be in the ballpark of $10^{-8}$ \cite{Glashow:2014iga}, which happens to be within reach at LHCb's run 2. An entirely analogous reasoning applies for the purely leptonic modes $B_s \to \ell_i^\pm \ell_j^\mp$. Similarly as eq. (\ref{eq:BKlilj}) one has
\be
\mc B (B_s \to \ell_i^\pm \ell_j^\mp) ~\simeq~ 
5\% \, \cdot \, \mc B (B_s \to \mu^+ \mu^-) \, \cdot \,
\frac{| (U^\ell_L)_{3i} |^2 | (U^\ell_L)_{3j} |^2}{|(U^\ell_L)_{32}|^4}~.
\ee
Therefore, purely leptonic LFV decays of the $B_s$ may well be within reach of LHCb during run 2, if the $U$-matrix factor on the r.h.s. is of order unity (or larger!) for at least one LFV mode.\footnote{%
For a (rough) comparison, we should keep in mind that at run 2 the LHCb is expected \cite{Bediaga:2012py} to provide a first measurement of $\mc B (B_d \to \mu^+ \mu^-)$, which in the SM is about 3\% of $\mc B (B_s \to \mu^+ \mu^-)$.}

From the previous line of argument about the $U_L^\ell$-matrix suppression, one may expect the $B_s \to e \mu$ to be the most difficult to access among the purely leptonic LFV modes. It is therefore useful to search for additional decays, that can give access to the same physics, while being comparably (or, hopefully, more) accessible experimentally. 

Here we would like to point out that, in the $B_s \to \mu e$ channel, one such `proxy' decay is provided by the inclusion of an additional hard photon in the final state. In fact, the additional photon replaces the chiral-suppression factor, of order ${\rm max}(m_{\ell_1},m_{\ell_2})^2 / m_{B_s}^2$, with a factor of order $\alpha_{\rm em} / \pi$. In the case of the $\mu e$ channel these two factors are respectively $4 \times 10^{-4}$ and 2 per mil. The actual enhancement of $\mc B(B_s \to \mu e \gamma)$ over the non-radiative counterpart needs to be worked out by explicit calculation.

We thus compute the decay of a generic pseudo-scalar meson to $\ell_{1}^{+} \ell_{2}^{-} \gamma$, and study in detail the cases of $B_s, B_d, K$ as initial state and of $\mu e$ as final state. (To fix notation, formulae are given for $B_q$.) The photon energy $E_\gamma$ is integrated above an experimental cut around 60 MeV, comparable to the experimental resolution on the total invariant mass of the final states, which has to yield the decaying-meson mass. We subsequently compare the resulting radiative rates with the non-radiative ones. We find an O(1) factor for the $B_q$ cases and at the percent level for the kaon case. Therefore, if experimental efficiencies for the radiative vs. the non-radiative $B_q$ decays are comparable, measurements of the radiative counterparts of LFV decays will provide crucial quantitative tests of the new-physics scenario responsible for a possible LFV signal.

\bigskip

\begin{figure}[t]
\begin{center}
  \includegraphics[width=0.30\linewidth]{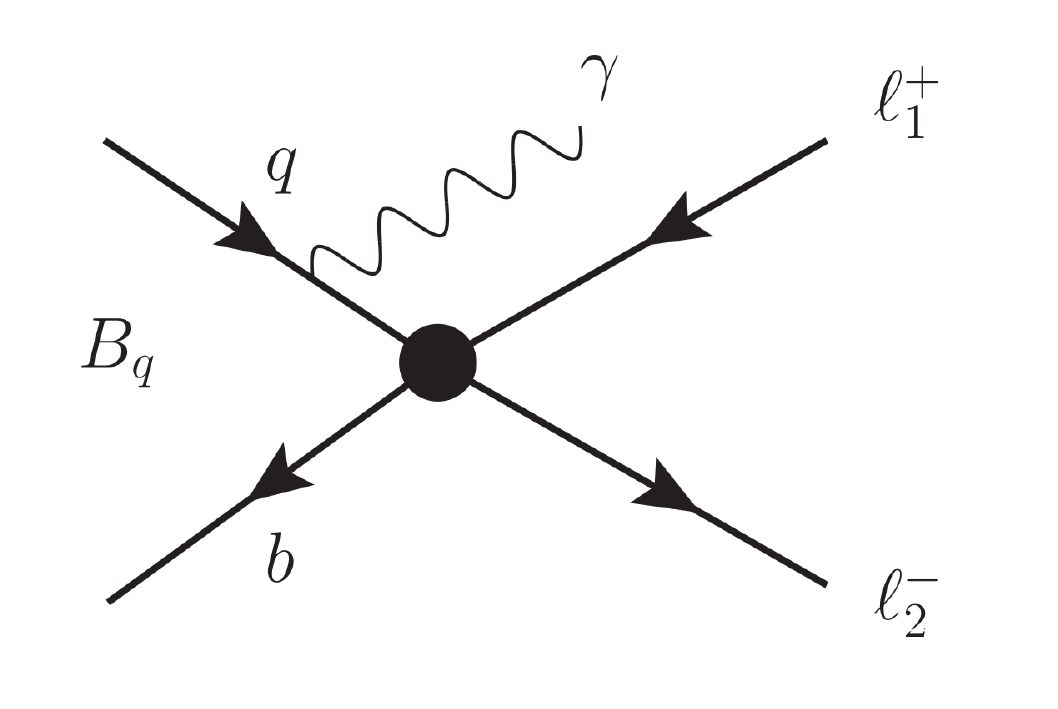} \hspace{1cm}
  \includegraphics[width=0.30\linewidth]{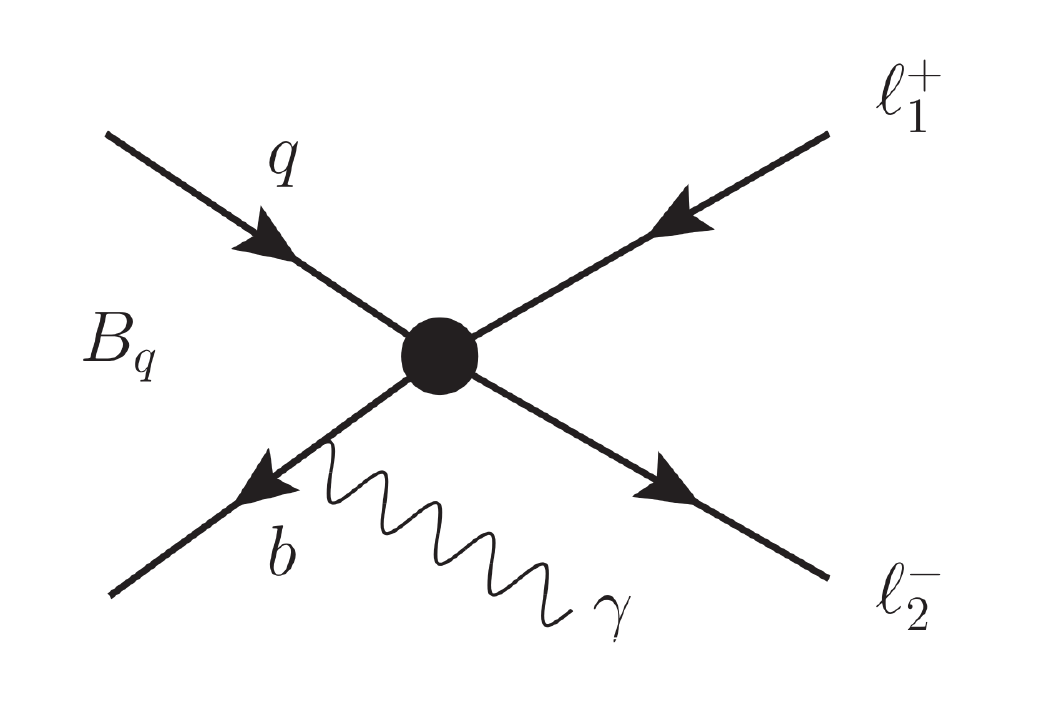}
  
  \includegraphics[width=0.30\linewidth]{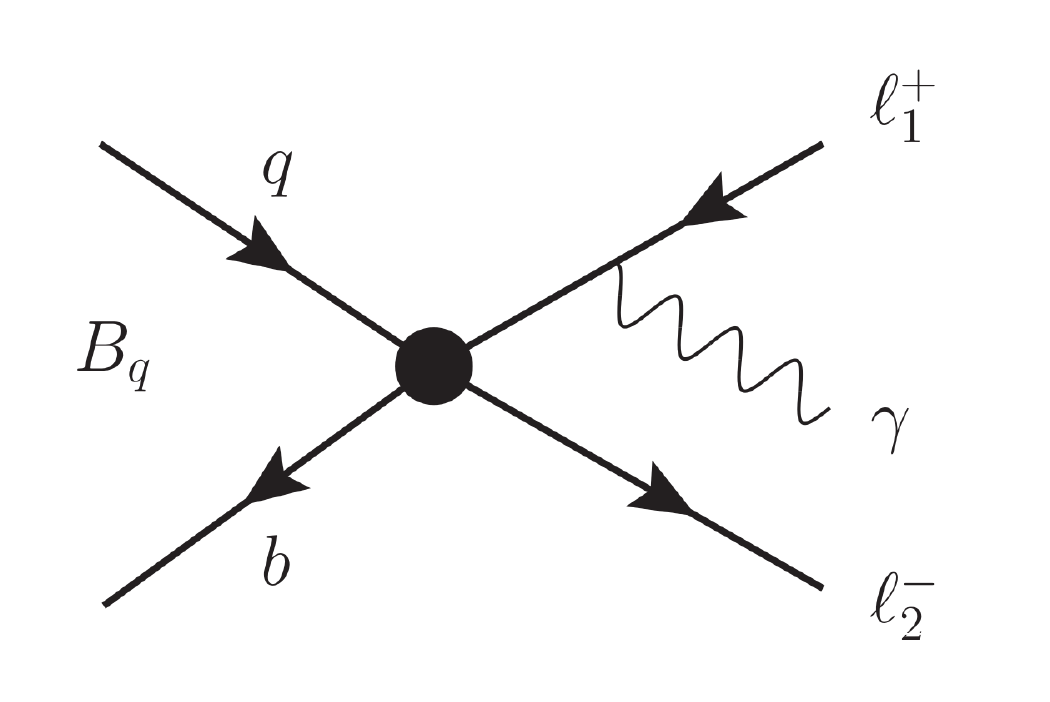} \hspace{1cm}
  \includegraphics[width=0.30\linewidth]{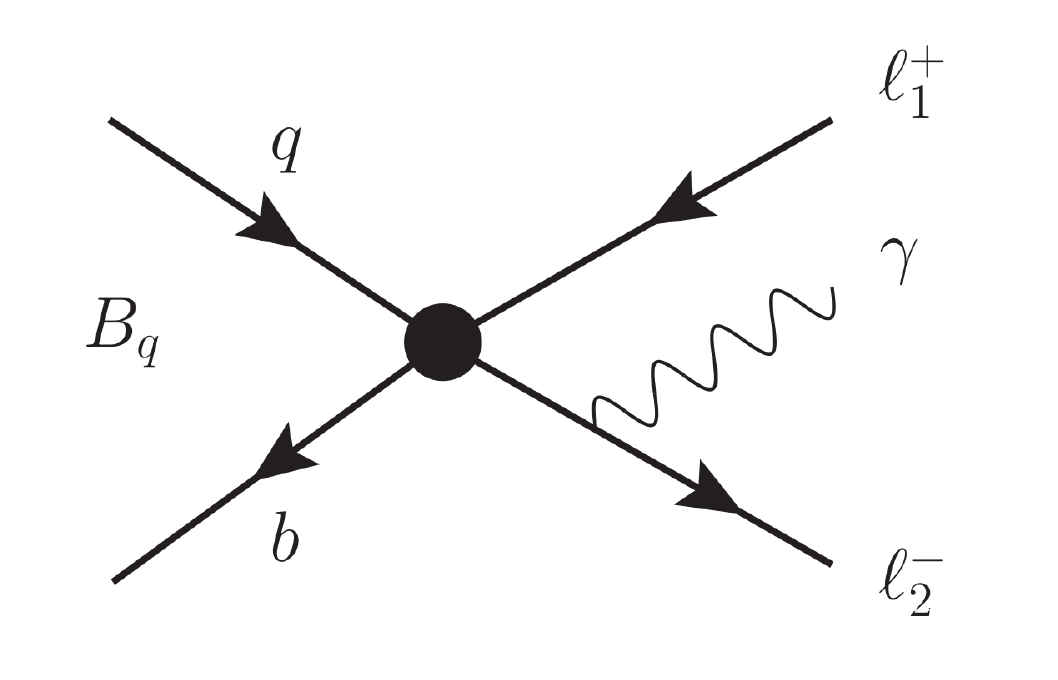}
  \caption{Diagrams contributing to $B_q \to \bar \ell_1 \ell_2 \gamma$, within the interaction in eq. (\ref{eq:HNP}). The black dot denotes the insertion of $\mc O_9$ or $\mc O_{10}$.}
  \label{fig:diags}
\end{center}
\end{figure}

\noindent {\bf Observables --} Within the interaction in eq. (\ref{eq:HNP}) the contributions to the process $B_{q} \rightarrow \ell_{1}^{+} \ell_{2}^{-} \gamma$ are given by the diagrams in fig. \ref{fig:diags}, where the black dot denotes the insertion of the (LFV counterparts of the) operators $\mc O_{9}$ or $\mc O_{10}$, in the following normalization \cite{Bobeth:2001sq}
\bea
\mc O_9 &=& \frac{\alpha}{4 \pi} \, (\bar b_L \gamma^\lambda q_L) \, 
(\bar \ell_2  \gamma_\lambda \ell_1)~,\nn \\
\mc O_{10} &=& \frac{\alpha}{4 \pi} \, (\bar b_L \gamma^\lambda q_L) \, 
(\bar \ell_2  \gamma_\lambda \gamma_5 \ell_1)~.
\eea
\noindent The diagrams in fig. \ref{fig:diags} correspond to two amplitudes: the one where the photon is emitted from the $B$-meson quarks, denoted as $A^{(1)}$, and the one due to Bremsstrahlung from the final-state leptons, denoted as $A^{(2)}$. Accordingly, the decay width is the sum of three terms, those coming from the magnitudes squared of the above amplitudes, and that  due to the interference between $A^{(1)}$ and $A^{(2)}$. In a notation close to ref. \cite{Melikhov:2004mk}, these three contributions read:
\bea
\label{eq:dGa1}
\frac{d^{2} \Gamma^{(1)}}{d \hat{s} d\hat{t}} &=& \frac{G_{F}^{2} \alem^{3} M_{B_{q}}^{5}}%
{2^{10} \pi^{4}} |V_{tb}^{{\phantom{*}}} V_{tq}^{*}|^{2} \left[ 
x^{2} B_{0}^{(1)} + x \, \xi B_{1}^{(1)} + \xi^{2} B_{2}^{(1)}
\right]~,\\
[0.2cm]
\label{eq:dGa2}
\frac{d^{2} \Gamma^{(2)}}{d \hat{s} d\hat{t}} &=& \frac{G_{F}^{2} \alem^{3} M_{B_{q}}^{5}}%
{2^{10} \pi^{4}} |V_{tb}^{{\phantom{*}}} V_{tq}^{*}|^{2} 
\left( \frac{f_{B_{q}}}{M_{B_{q}}} \right)^{2} 
\frac{1}{(\th - \mh_{2}^{2})^{2} (\uh - \mh_{1}^{2})^{2}} \times \nn \\
&&\left[
x^{2} B_{0}^{(2)} + x \, \xi B_{1}^{(2)} + \xi^{2} B_{2}^{(2)}
\right]~,\\
[0.2cm]
\label{eq:dGa12}
\frac{d^{2} \Gamma^{(12)}}{d \hat{s} d\hat{t}} &=& \frac{G_{F}^{2} \alem^{3} M_{B_{q}}^{5}}%
{2^{10} \pi^{4}} |V_{tb}^{{\phantom{*}}} V_{tq}^{*}|^{2} 
\left( \frac{f_{B_{q}}}{M_{B_{q}}} \right) 
\frac{1}{(\th - \mh_{2}^{2})(\uh - \mh_{1}^{2})} \times \nn \\
&&\left[
x^{2} B_{0}^{(12)} + x \, \xi B_{1}^{(12)} + \xi^{2} B_{2}^{(12)}
\right]~,
\eea
where
\be
\sh = \frac{(p - k)^{2}}{M_{B_{q}}^{2}}~,~~~
\th = \frac{(p - p_{1})^{2}}{M_{B_{q}}^{2}}~,~~~
\uh = \frac{(p - p_{2})^{2}}{M_{B_{q}}^{2}}~,
\ee
and the $\sh, \th, \uh$ variables fulfill the constraint
\be
\sh + \th + \uh = 1 + \mh_{1}^{2} + \mh_{2}^{2}~.
\ee
Here $p, k, p_1$ and $p_2$ denote the momenta of, respectively, the initial meson, the emitted photon, and the leptons $\ell_{1,2}$, whose masses are $m_{1,2}$. The hat denotes that the given variable has been made dimensionless by normalizing
it to an appropriate power of $M_{B_{q}}$. Furthermore
\be
x ~\equiv~ 1 - \sh~,~~~ 
\xi ~\equiv~ \uh - \th + \frac{\mh_{2}^{2} - \mh_{1}^{2}}{\sh}~.
\ee
The $B_{i}^{(j)}$ functions are defined as follows:
\bea
B_{0}^{(1)} &=& \left( F_{V}^{2}(\sh) + F_{A}^{2} (\sh) \right)
\left[ \left(\sh - \frac{\Mh^{2} \mh^{2}}{\sh} \right)
\left( \left| C_{9} \right|^{2} + \left| C_{10} \right|^{2} \right) +
4 \mh_{1} \mh_{2} 
\left( \left| C_{9} \right|^{2} - \left| C_{10} \right|^{2} \right)
\right]~,
\nn \\
B_{1}^{(1)} &=& 8 \sh \, F_{V}(\sh) F_{A}(\sh) 
\re \left( C_{9}^{\phantom{*}} C_{10}^{*} \right)~,\nn \\
[0.2cm]
B_{2}^{(1)} &=& \sh 
\left( F_{V}^{2}(\sh) + F_{A}^{2} (\sh) \right)
\left( \left| C_{9} \right|^{2} + \left| C_{10} \right|^{2} \right)~,
\eea

\bea
B_{0}^{(2)} &=& 2 \Mh^{2} 
\left( 2 \sh \, \rho \, (1 - \mh^{2} ) + 
x^{2} \left( 1 - \frac{\Mh^{2} \mh^{2}}{\sh^{2}} \right)
\right) |C_{10}|^{2} + \nn \\
&& 2 \mh^{2} 
\left( 2 \sh \, \rho \, (1 - \Mh^{2} ) + 
x^{2} \left( 1 - \frac{\Mh^{2} \mh^{2}}{\sh^{2}} \right)
\right) |C_{9}|^{2}~,\nn \\
B_{1}^{(2)} &=& \frac{4 x^{2} \Mh \mh}{\sh} \left[ \Mh^{2} |C_{10}|^{2} + 
\mh^{2} |C_{9}|^{2} \right]~,\nn \\
[0.2cm]
B_{2}^{(2)} &=& 2 \Mh^{2} 
\left(2 \sh (\mh^{2} -1 ) - x^{2} \right)|C_{10}|^{2} + 2 \mh^{2} 
\left(2 \sh (\Mh^{2} -1 ) - x^{2} \right)|C_{9}|^{2}~,
\eea

\bea
B_{0}^{(12)} &=& - 8 x \, F_{V} (\sh)
\left( \mh_{1}^{2} + \mh_{2}^{2} - \frac{\Mh^{2} \mh^{2}}{\sh} \right)
\re \left( C_{9}^{\phantom{*}} C_{10}^{*} \right) + \nn \\
&& \frac{4 \Mh \mh}{\sh} \, F_{A} (\sh) \left[( \Mh^{2} - \sh )
(1 - \mh^{2} ) |C_{10}|^{2} + ( \mh^{2} - \sh )
(1 - \Mh^{2} ) |C_{9}|^{2} \right] ~,\nn \\
B_{1}^{(12)} &=& - 8 x \Mh \mh \, F_{V} (\sh)
\re \left( C_{9}^{\phantom{*}} C_{10}^{*} \right)
- 4 x F_{A} (\sh) \left[ \Mh^{2} |C_{10}|^{2} + 
\mh^{2} |C_{9}|^{2} \right]~,\nn \\
[0.2cm]
B_{2}^{(12)} &=& 4 \sh \Mh \mh \, F_{A} (\sh)
\left( |C_{10}|^{2} + |C_{9}|^{2} \right)~,
\eea
where
\be
\Mh = \mh_{1} + \mh_{2}~,~~~
\mh = \mh_{2} - \mh_{1}~,~~~
\rho = \frac{(\sh - \Mh^{2})(\sh - \mh^{2})}{\sh^{2}}~.
\ee
In the lepton-flavor conserving limit, these equations reproduce the results of ref. \cite{Melikhov:2004mk}, apart from an overall sign typo in eqs. (2.13) and (3.3) of that paper.

The $f_{B_{q}}$ decay constant and the $B_{q} \rightarrow \gamma$ form factors involved
are defined by \cite{Kruger:2002gf}
\bea
\label{eq:fB}
\< 0 | \, \bar{b} \, \gamma^{\mu} \gamma_{5} \, q \, | B_{q} (p) \> &=& i p^{\mu} f_{B_{q}}~,\\
\label{eq:FV}
\< \gamma(k,\epsilon) | \, \bar{b} \gamma^{\mu} q \, | B_{q} (p) \> &=& e \, \epsilon^{*}_{\nu} \epsilon^{\mu \nu \rho \sigma} p_{\rho} k_{\sigma} \, 
\frac{F^{(B_q)}_{V}(\sh)}{M_{B_{q}}}~,\\
\label{eq:FA}
\< \gamma(k,\epsilon) | \, \bar{b} \gamma^{\mu} \gamma_{5} q \, | B_{q} (p) \> &=& i e \, \epsilon^{*}_{\nu} \, (g^{\mu \nu} pk - p^{\nu} k^{\mu}) \, \frac{F^{(B_q)}_{A}(\sh)}{M_{B_{q}}}~.
\eea

\noindent For the $F_{V,A}^{(B_q)}$ form factors we use the parameterization in \cite{Melikhov:2004mk}. (Within few percent, the $B_s \to \gamma$ form factors coincide with the $B_d \to \gamma$ ones. Such differences are clearly negligible in our context.)

For the $K \to \gamma$ form factors, defined in complete analogy to eqs. (\ref{eq:FV})-(\ref{eq:FA}) and denoted as $F_{V,A}^{(K)}$, we instead adapt to light quarks the recent analysis \cite{Kozachuk:2015kos} of heavy-meson transition form factors. The latter is based on the relativistic constituent quark model \cite{Anisovich:1996hh,Melikhov:2001zv}. This model makes no fundamental difference between heavy and light mesons, and as such it is applicable to either case. For heavy quarks, the analytic expressions for the form factors from the constituent quark model reproduce the known results from QCD for heavy-to-heavy and heavy-to-light form factors. For light quarks, properties known for QCD in the chiral limit constrain the structure of the axial current of the constituent quarks \cite{Lucha:2006rq}. Form-factor predictions within this model are thereby expected to be within about 10\% \footnote{Such accuracy is more than sufficient in our case, given that the $K \to \mu e \gamma$ mode will turn out to be less interesting than the $B$-decay counterparts, see eq. (\ref{eq:RoNR}) below.} with respect to ones based on first-principle approaches.\footnote{One such approach is to relate the $K \to \ell \ell' \gamma$ form factors to those of $K \to \ell \nu  \gamma$ \cite{Cirigliano:2011ny} using isospin symmetry. We thank the referee for a useful remark in this context.} We calculated the $K \to \gamma$ form factors making use of the model parameters from ref. \cite{Melikhov:2000yu}. The calculation may be parameterized by the simple formula
\be
\label{eq:ff_K}
F_{V,A}^{(K)}(\sh) ~=~ \frac{Q_d F_{V,A}^{(d)}(0) \pm Q_s F_{V,A}^{(s)}(0)}{1 - 
\sh \left( \frac{M_K}{M_{V,A}} \right)^2}~,
\ee
with $Q_{d,s} = -1/3$, $F^{(d,s)}_V(0)= \{-0.216,\,-0.18\}$, $F^{(d,s)}_A(0)=\{0.201,\,0.19\}$ and $M_{V,A} = 0.89$ GeV. For the meson decay constants we use $f_{B_{d,s}} = \{ 0.186, 0.224 \}$ GeV \cite{Dowdall:2013tga} and $f_K = 0.155$ \cite{Carrasco:2014poa}. The rest of the relevant input parameters are taken from \cite{Agashe:2014kda}.

The branching ratio for the corresponding non-radiative decay $B_{q} \rightarrow \ell_{1}^{+} \ell_{2}^{-}$ reads
\bea \label{eq:NR}
\mathcal{B} (B_{q} \rightarrow \ell_{1}^{+} \ell_{2}^{-}) &=&
\tau_{B_{q}} \frac{G_{F}^{2} \alem^{2} M^{3}_{B_{q}} f^{2}_{B_{q}}}{2^{6} \pi^{3}}
|V_{tb}^{{\phantom{*}}} V_{tq}^{*}|^{2} \sqrt{\rho} \times \\ \nn
&& \left[
(1 - \mh^{2} )|F_{P} + \Mh C_{10}|^{2} +
(1 - \Mh^{2} )|F_{S} - \mh C_{9}|^{2}
\right]~,
\eea
with
\be
F_{S, P} = M_{B_{q}} \,
\frac{m_{b} C_{S, P} -  m_{q} C'_{S, P}}{m_{b} + m_{q}}~.
\ee
In the lepton-flavor conserving limit this formula reproduces exactly the corresponding one in ref. \cite{Bobeth:2001sq}.

We note that the above branching-ratio formulae refer to `instantaneous' $B_q$ decays. This observation is relevant for a precision calculation of $B_s$ decay branching ratios. In this case, one should replace $\tau_{B_s}$ with $\tau_{B_H}$ (where $B_H$ is the heavier of the $B_s - \bar B_s$ mass eigenstates), to account for the large width difference in the $B_s$ system \cite{DeBruyn:2012wk}.

Furthermore, as already remarked, in our numerics we integrate the photon energy spectrum above $E_{\rm cut} = 60$ MeV, comparable to the experimental resolution on the total invariant mass of the final states, which has to yield the decaying-meson mass.\footnote{We take $E_{\rm cut} = 60$ MeV from the $B_s \to \mu \mu$ case \cite{Buras:2012ru}. When including an additional photon, the correct value to be taken for $E_{\rm cut}$ may be slightly higher \cite{JFMarchand}. However, the choice of its precise value has a minor impact on our predictions, as $E_{\rm cut}$ only modifies the $\sh$ integration endpoint $\sh_{\rm max} = 1 - 2 E_{\rm cut}/M_{B_q}$.} The above cut doesn't completely exclude `soft' photons, namely ones such that $E_\gamma \ll m_{B_q}/2$. Their effect can be summed to all orders in the soft-photon approximation, and leads to a multiplicative correction factor to the non-radiative rate, of the order of 10\%, as discussed in \cite{Buras:2012ru}. However, within LHCb, the effect of soft final-state radiation is corrected for by an appropriate Monte Carlo.

\bigskip

\noindent {\bf Numerical Analysis --} One can now compute the predictions for the radiative and non-radiative cases within the interaction in eq.(\ref{eq:HNP}) \cite{Glashow:2014iga}, whereby the shifts to $C_9$ and $C_{10}$ read
\be
\label{eq:deltaC9}
\delta C_9 = - \delta C_{10} = \frac{G}{2} \,
\frac{(U^{d*}_{L})_{33} (U^{d}_{L})_{3q} (U^{\ell*}_{L})_{3 \ell_2} (U^{\ell}_{L})_{3 \ell_1}}{-\frac{4 G_F}{\sqrt 2} V^*_{tb} V^{\phantom{*}}_{tq} \frac{\alem (m_b)}{4 \pi}} ~.
\ee
These predictions will depend on two basic parameters, the overall strength $G$ of the interaction in eq. (\ref{eq:HNP}), and the product of four chiral rotations turning the fermion fields $(\bar b' b') (\bar \tau' \tau')$ into the fields relevant for the process, $(\bar b q) (\bar \ell_2 \ell_1)$. This product of four $U$-matrix entries will be denoted as $U_4$ for brevity.

The parameters $G$ and $U_4$ are completely unknown and we have at best some guiding criteria to fix them to reasonable ranges:
\begin{itemize}

\item Since, for a given process, $G$ and $U_4$ always appear as a product, it is always possible to shuffle an arbitrary numerical factor between $G$ and $U_4$. As a consequence, to fix a reasonable range for $G$ with any confidence, one may consider predictive models for $U_4$, as in ref. \cite{Guadagnoli:2015nra} (see also \cite{Boucenna:2015raa}). One obtains new-physics scales $\Lambda_{NP} = 1/\sqrt{G}$ between 750 GeV and 5 TeV.\footnote{As emphasized in \cite{Guadagnoli:2015nra}, these mass scales may appear low for, say, a $Z'$ as the underlying mediator of the interaction (\ref{eq:HNP}). However, it must be remembered that this interaction couples primarily to the third generation.} We then assume $4\times10^{-8}~{\rm GeV}^{-2}\leq G \leq 2\times10^{-6}~{\rm GeV}^{-2}$.

\item The PMNS-matrix anarchy suggests that the leptonic part of $U_4$ may be of O(1). As concerns the $U^d_L$ matrix entries $33$ and $3q$, one can assume them to be close in magnitude to the CKM entries $V_{tb}$ and $V_{tq}$ respectively.\footnote{%
This assumption should actually hold to a good extent, provided that new interactions other than eq. (\ref{eq:HNP}) are indeed negligible, as assumed here.} We therefore consider the range $10^{-4} \leq U_4 \leq 0.05$,  keeping in mind that $B_s$ would correspond to $U_4 \sim |V_{ts}| \simeq 0.04$ and $B_d$ to $U_4 \sim |V_{td}| \simeq 0.008$.

\end{itemize}

\begin{figure}[t!]
\begin{center}
  \includegraphics[width=0.65\linewidth]{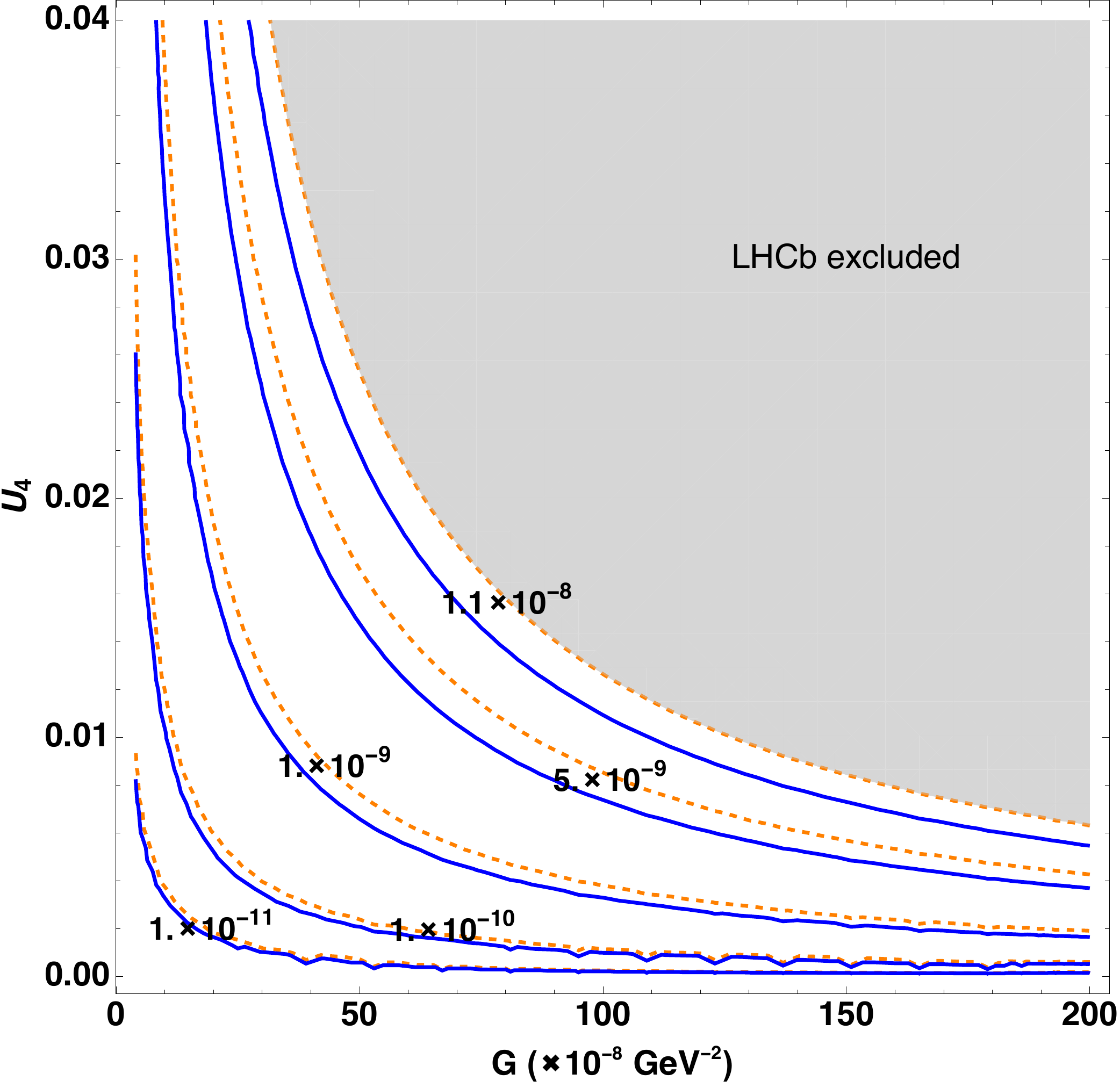} \hfill
  \caption{$\mc B(B_{s} \rightarrow e^{\pm} \mu^{\mp} \gamma)$ (blue, solid) and $\mc B(B_{s} \rightarrow e^{\pm} \mu^{\mp})$ (orange, dashed) as a function of $G$ vs. the product of $U$-matrix entries appearing in eq. (\ref{eq:deltaC9}), and denoted as $U_4$. (See text for more details.) The gray area is excluded by the LHCb upper limit on the non-radiative decay \cite{Aaij:2013cby}.}
  \label{fig:plot}
\end{center}
\end{figure}

To get an idea of the resulting predictions, we note that the upper limit $\mc B(B_{s} \rightarrow \mu^\pm e^\mp) < 1.1 \times 10^{-8}$ from LHCb \cite{Aaij:2013cby} corresponds to $G \times U_4 = 1.3 \times 10^{-8}~{\rm GeV}^{-2}$, which is about 6 times smaller than the product of our highest allowed values for $G$ and $U_4$. A general picture of the predictions for the $B_{s} \to \mu e \gamma$ branching ratio and its non-radiative counterpart as a function of $G$ vs. $U_4$ in the above-mentioned ranges is presented in fig. \ref{fig:plot}. The gray area denotes the parameter space excluded by the LHCb $B_s \to \mu e$ search of ref. \cite{Aaij:2013cby}.

The figure shows that the radiative mode is slightly enhanced with respect to the non-radiative counterpart. Actually, within our considered model, where the shifts to $C_9$ and $C_{10}$ differ only by a sign, the $|G \times U_4|^2$ dependence cancels altogether in the radiative over non-radiative ratio, and we find
\be
\label{eq:RoNR}
  \frac{\mc B(B_{s} \rightarrow \mu e \gamma)}{\mc B(B_{s} \rightarrow \mu e)} = 1.3 ~,~~~
  \frac{\mc B(B_{d} \rightarrow \mu e \gamma)}{\mc B(B_{d} \rightarrow \mu e)} = 1.2~,~~~
  \frac{\mc B(K \rightarrow \mu e \gamma)}{\mc B(K \rightarrow \mu e)} = 2.7 \times 10^{-2} ~.
\ee
As a consequence, assuming experimental efficiencies for radiative and non-radiative cases to be comparable, the measurement of the $B_q$ radiative decay along with the non-radiative one offers a precious cross-check of the new-physics mechanism responsible for a possible LFV signal. In the $K$ case instead, the radiative mode is too suppressed to be potentially interesting, unless the $K \to \mu e$ mode is found at an unexpectedly large rate.

The numbers in eq. (\ref{eq:RoNR}) can be intuitively understood as follows. First note that the Bremsstrahlung contribution $d \Gamma^{(2)}$ to the radiative decay comes with a factor of $(f / M)^2$, with $M$ the mass of the decaying meson and $f$ its decay constant, as well as with a chiral suppression factor. On the other hand, both of these suppression factors are absent in the direct-emission contribution $d \Gamma^{(1)}$. 
Therefore, the radiative decay will be competitive with the non-radiative one -- the latter also $(f/M)^2$ as well as chirally suppressed -- to the extent that the direct-emission contribution can dominate, which occurs whenever $f/M$ is small enough, that is the case for both $B_d$ and $B_s$, but not for kaons. In other words, the larger the ratio $M / f$, with $M$ the mass of the decaying meson and $f$ its decay constant, the larger the parametric enhancement of the radiative decay over the non-radiative counterpart.

\bigskip

\noindent {\bf Conclusions --} We studied in detail the prediction for the LFV decays of a $K, B_d$ or $B_s \to \mu^\pm e^\mp \gamma$. These decays are `proxies' to the corresponding non-radiative decays, in that LFV physics of the kind discussed in the text will produce both sets of decays, the chiral-suppression factor in the non-radiative modes being replaced by a factor of the order of $\alpha / \pi$ in the radiative ones. We found that predictions for the total branching ratios for $B_{d,s} \to \mu^\pm e^\mp \gamma$ exceed by about 30\% those for the corresponding non-radiative modes. Taking into account that experimental efficiencies may be slightly lower for $\mu e \gamma$ than they are for just $\mu e$ pairs, inclusion of the proposed radiative modes realistically corresponds to a {\em doubling} of the statistics as compared to the purely non-radiative modes, for either of $B_d$ and $B_s$.

\section*{Acknowledgements}
The work of DG is partially supported by the CNRS grant PICS07229. The authors acknowledge K. Lane for discussions on related subjects, and J.-F. Marchand for remarks on the manuscript.


\providecommand{\href}[2]{#2}\begingroup\raggedright\endgroup

\end{document}